\documentclass{ws-ijbc}
\usepackage{color}
\usepackage{epsfig}  
\usepackage{graphicx}

\begin{document}

\catchline{}{}{}{}{} 

\markboth{Prignano \em{et al.}}{Synchronization of moving integrate and fire oscillators}

\title{SYNCHRONIZATION OF MOVING INTEGRATE AND FIRE OSCILLATORS}

\author{Luce Prignano}

\address{Departament de Fisica Fonamental, Universitat de Barcelona, Marti i Franques 1\\
Barcelona, 08028, Spain\\
luce@ffn.ub.es\footnote{Departament de Fisica Fonamental, Universitat de Barcelona, Marti i Franques 1,
Barcelona, 08028, Spain}}

\author{Oleguer Sagarra}
\address{Departament de Fisica Fonamental, Universitat de Barcelona, Marti i Franques 1\\
Barcelona, 08028, Spain\\
osagarra@ffn.ub.edu}

\author{Pablo M. Gleiser}
\address{Centro At\'omico Bariloche, Instituto Balseiro,
8400 R\'{\i}o Negro, Argentina \\
gleiser@cab.cnea.gov.ar
}

\author{Albert Diaz-Guilera}
\address{Departament de Fisica Fonamental, Universitat de Barcelona, Marti i Franques 1\\
Barcelona, 08028, Spain\\
albert.diaz@ub.edu}

\maketitle

\begin{history}
\received{(to be inserted by publisher)}
\end{history}

\begin{abstract}
We present a model of integrate and fire oscillators that move on a plane. The phase of the oscillators evolves linearly in time and when it reaches a threshold value they fire choosing their neighbors according to a certain interaction range.
Depending on the velocity of the ballistic motion and the average number of neighbors each oscillator fires to, we identify different regimes shown in a phase diagram. We characterize these regimes by means of novel parameters as the accumulated number of contacted neighbors. 

\end{abstract}


\section{Introduction}
\noindent 
Among the many emerging phenomena we observe in nature, synchronization is one of the most paradigmatic examples. It can be roughly understood as the collective dynamics of units whose internal state evolves periodically in time and when they interact tend to synchronize their internal variables \cite{prk01}. The achievement of the final synchronized state (if any) strongly depends on the interaction pattern of the system \cite{blmch06,adkmz08}.
Up to now, synchronization has been mainly analyzed in fixed topologies but we are witnessing the first evidences that links between agents can evolve in time \cite{ofm07,onnela07,vmdc08}. 
The case in which this evolution of the network topology is an effect of the agents mobility is a particularly interesting case \cite{bffr06,tjp03,bschdms06}. 
The effect of this changing patterns of interaction on synchronization features has been analyzed in different settings, for instance in chemotaxis \cite{t07}, mobile ad hoc networks
\cite{r01}, wireless sensor
networks \cite{sy04}, and the expression of segmentation clock genes
\cite{umi10}.

In the recent literature, studies on synchronization in dynamically evolving complex networks have been mainly concentrated in the case when the topology changes very fast.
This is the so-called fast-switching approximation (FSA) {\cite{bbh04b,frasca08,psbs06,sbr06}}, which replaces the real interaction between agents by the "mean field assumption" that all agents interact with an effective strength that corresponds to the probability that any pair of agents are connected.

Recently, it has been proposed a general framework of mobile oscillator networks
where agents perform random walks in a two-dimensional (2D) plane \cite{fkd11}.
It has been shown  that FSA fails when the time scale of local synchronization
is shorter than the time scale of the topology change due to the agent
motion. 
New behaviors arise due to the interplay between 
instantaneous network topology, agent motion,
and interaction rules. This framework, that reduces to FSA when velocity is high enough, is valid for models whose evolution can be well approximated by linear dynamics. This actually holds for models such as populations of Kuramoto oscillators \cite{kuramoto84,abprs05},  whose evolution, after a short transient time, is very well described by a set of linear equations that can be solved in terms of spectral properties of the Laplacian matrix \cite{fkd11b}.

In the present paper, we focus on a dynamical system, a population of integrate and fire oscillators (IFO), where linearization is not a good approximation, since the evolution takes place in two different time scales. One for the slow evolution of the internal state variables (the phase and the orientation) and the other for the fast interaction between the units (pulse coupling). 
During the last years it has been shown that the  interaction structure plays a fundamental role in the dynamics of IFO networks.
\citet{zbh09} observed different dynamical regimes due to network connectivity in
a system formed by inhibitory integrate-and-fire neurons that were randomly connected.
Also, the underlying network structure can affect the speed with which the system reaches the synchronized state, as studied by \citet{ggt11}. Usually, IFO have been used to model neural systems but we can also find some examples of applications in other fields, as for example in economy \cite{edga11}.
Models where the oscillators do not remain fixed, and the network of interactions changes with time can find a direct
application in biological systems such as flashing fireflies, that interchange light signals while searching for potential
mates~\cite{ms90,rdgwk11}.

In the present case we will show that the interplay between agents motion and phase evolution towards a synchronized state presents different asymptotic behaviors, reminiscent of the observation in Kuramoto oscillators \cite{fkd11} and agents using communication protocols \cite{bd11}. We identify, furthermore, the possible mechanisms in the different regions of the parameter space.

The organization of the paper is as follows. In the next section we introduce the model. Then we show the results for different regions of the parameter space,  velocity of the agents and range of interaction, and later we identify the different microscopic mechanisms that lead the system to a globally synchronized state.

\section{The model}

We propose a setting in which a population of $N$ integrate and fire oscillators (IFOS) \cite{ms90} move at a constant velocity $V$ in a bidimensional plane of size $L$ with periodic boundary conditions. Each agent has two degrees of freedom corresponding to an internal phase $\phi \in [0,1]$ and orientation $\theta_i \in [0,2\pi]$, both randomly set in an uniform manner at the initial configuration.

The evolution of the system takes place on two different timescales. 
The slow timescale sets the pace at which the phases of the agents increase uniformly with period $\tau$,
\begin{equation}
\frac{d\phi_i}{dt}=\frac{1}{\tau}
\end{equation}
until they reach a maximum value of 1, when a firing event occurs. Then the phase is reset and the oscillator is randomly reoriented. 
Upon this event at time $t$, the firing oscillator 
influences its \emph{nearest neighbors} (oscillators at minimal distance) producing an update in their phases by a factor $\epsilon$:
\begin{equation}
\phi_{i}(t^-)=1 \Rightarrow \left \{ \begin{array}{l} \phi_i(t^+)=0\\ \phi_{\text{nn}}(t^+)=(1+\epsilon)\phi_{\text{nn}}(t^-) \\ \theta_i(t^+) \in  [0,2\pi]\end{array} \right. .
\end{equation}
The phase and orientation resetting corresponds then to the fast time scale.
If the neighbor's phase update overcomes the phase maximum, another firing event is triggered and this process goes on repeatedly until all shots have ceased. At this point, the time $t$ runs again until the next firing event occurs. The system is synchronized when we encounter in the system a succession of consecutive firing events (avalanche) equal to the system size $N$, since after this fact all oscillators will remain synchronized forever because all of them will have the same period $\tau$ with or without interactions. For the sake of clarity we define the (discrete) time $T$, as the number of times a given oscillator (that we will identify with oscillator 1 in our computer simulations) has fired. This allows us to define $T_{sync}$ as the minimum number of (integer) cycles this reference oscillator takes to enter the synchronized state (i.e. the number of updates needed for an avalanche of size $N$ to occur).
\begin{figure}[htbp]
\begin{center}
\includegraphics[scale=0.5]{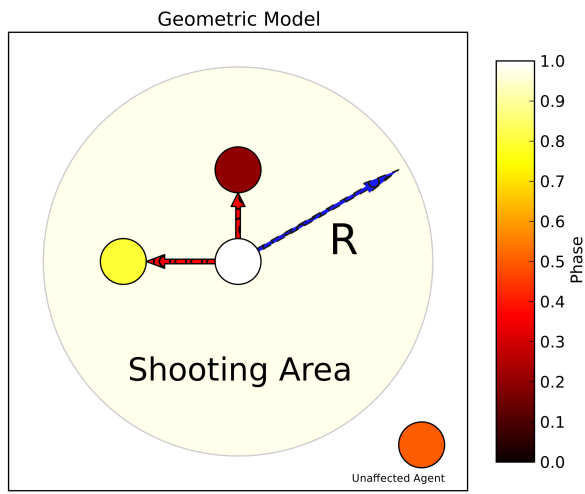}
\caption{The model of interaction between oscillators, based on geometrical constraints. 
Only those within a distance $R$ are affected by the firing of the central one.}
\label{fig1}
\end{center}
\end{figure}


We propose a geometric condition for neighbor selection upon a firing event as shown in figure \ref{fig1}: Every agent scans a circular area of radius $R$ around it and shots the neighbors therein. 
We introduce a parameter $r \in [0,1]$ that indicates the fraction of the system available for interaction and relates both $R$, $L$ variables and the average outgoing degree of the nodes of our evolving network
\begin{equation}
r=\frac{\pi R^2}{L^2} \quad \quad \langle k^{out} \rangle =(N-1)r.
\end{equation}
Throughout this paper, we have used fixed parameters $L=100$, $\tau=1$, $N=50$ and $\epsilon=0.02\sim \mathcal{O}(1/N)$, while analyzing the explicit dependence on the mobility parameters, $r$ and $V$.

Before proceeding to show the results of our simulations, we need to note that the type of proposed interaction range in this system has been reported to show statistical properties similar to a continuous percolation \cite{fkd11}, that in the case of static oscillators occurs for approximately $r_c\approx 4.51 / (N-1)=0.09$ \cite{dc02,bbw05}. In our range of study, we hope to observe some traces of this percolation as well as saturation properties observed in other moving oscillator systems at high speeds \cite{fkd11}. 

It is also important to notice that we have kept the dynamical evolution of the units at its maximal simplicity since we are mostly interested in the interplay between motion (and hence construction of a dynamical network) and internal dynamics and how synchronization emerges as a collective property of the system.

\section{Results}
\begin{figure}[htbp]
\begin{center}
\includegraphics[width=.95\columnwidth]{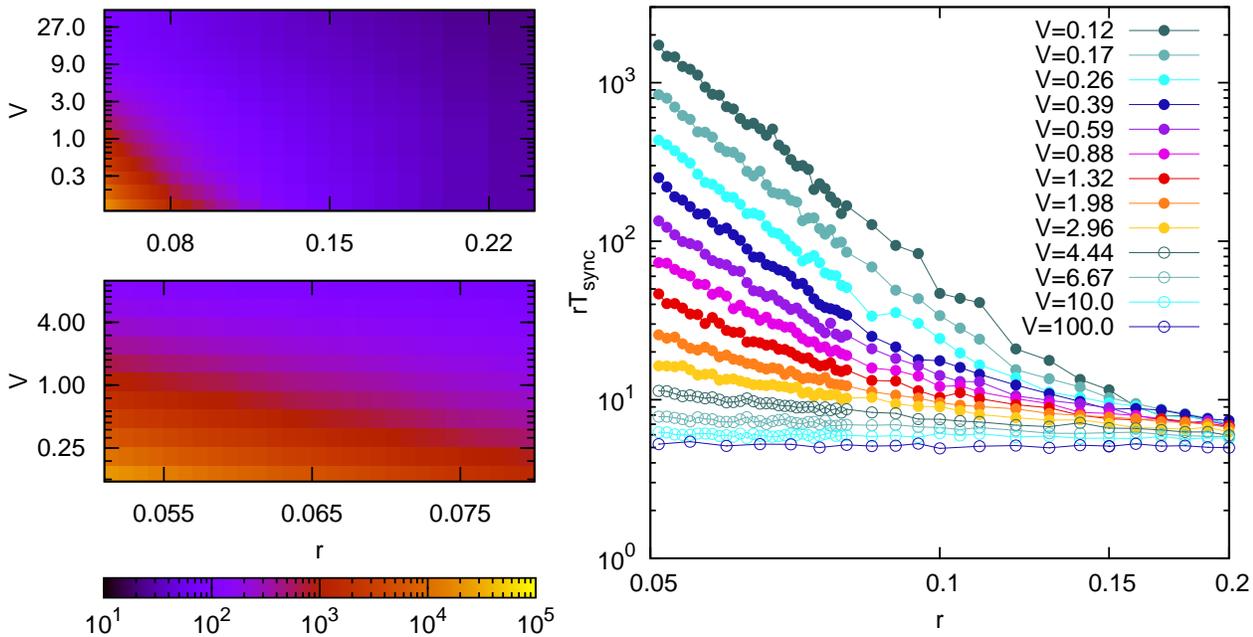}
\caption{(Colors online) On the left: Heat map of the synchronization time as a function of $r$ and $V$; in the top picture  a large region of the parameters space has been considered, while the bottom figure is a zoom on the most critical region ($r<0.08$, $V<0.25$). On the right: Profiles of the efficiency of the system $rT_{sync}$  against $r$, for several values of $V$.}
\label{fig2}
\end{center}
\end{figure}

We present in figure \ref{fig2} the results of our simulations. A preliminary observation points out that the roles and importances of $V$ and $r$ change throughout our map. 

For high enough values of $r$ ($r\gtrsim r_c$) , the synchronization time, $T_{sync}$, is almost unaffected by the values of the speed $V$. Although this time is dependent on $r$, 
its range of possible values is much narrower (by orders of magnitude) than below the value, $r_c$, that characterizes the static percolation transition. Actually, below this critical value the velocity plays the crucial role since for a fixed value of $r$ the synchronization time changes by a factor of $10^4$.

To bring further insight to the map, we show a profile of the efficiency of the system $rT_{sync}$ dependence on $r$ for various velocities. This efficiency variable balances the range of interactions (which has an "energy cost" related with the number of shots to different neighbors and their strength) and the number of shots that the system needs to synchronize (related with $T_{sync}$). 

It seems clear that mobility of agents helps to minimize energy consumption but as we approach the critical percolating value of $r_c$, we observe that a range of behaviors emerge. On one hand, we observe for high velocities that the efficiency of the process remains roughly constant independently of $r$, since the extreme mobility of the agents compensate the reduced range of its interactions and successfully diffuses the synchronization process around the system, a process that is equivalent to the observation in other settings \cite{frasca08,fkd11,bd11}. On the other hand, if the mobility of the agents is reduced, then the path to synchronization is more lengthly as well as more energy consuming. Synchronization is still possible (below  the percolation static limit $r_c$) but the time to achieve it grows very fast, even resulting in an effectively infinite time\footnote{In our simulations some realizations of the experiment did not reach synchronization, even working with a reduced number of oscillators and the described boundary conditions, fact that induces us to think in this direction.}. 

These results lead us to identify different regimes and the consequent transitions between them. On the one hand, we find a "topological" transition at the critical static value $r_c$, since above it there is basically no velocity dependence whereas below $r_c$ the influence of $V$ is determinant. On the other hand, below $r_c$, where synchronization is made possible by the mobility of the agents, we can identify a transition (depending on both $r$ and $V$) that separates two dynamical regimes.

\section{Mechanisms}

Up to this point, we have studied the efficiency of our system and detected several regimes strongly dependent on two factors, the mobility of our agents and its range of interaction. 

In this part of the paper we want to study the distinct mechanisms that the system uses in its path to synchronization. To this end, we introduce an order parameter  $\eta(T)=\langle\cos{(2\pi\phi(T))}\rangle$ that is an increasing function that measures the overall synchrony of our system, ranging from a uniform phase distribution of our oscillators ($\eta(T)=0$) to complete synchronization\footnote{Note that the average is calculated upon a firing event by the reference oscillator, hence our order parameter is an average of the phase difference of the other oscillators with respect to this one.} ($\eta(T)=1$). 

To couple our $V$ and $r$ control parameters we also introduce the number of distinct interactions per oscillator $N_c^i$ (accumulated encounters with different agents by a oscillator $i$). This value is bounded (on average) between a minimal starting value of $\langle N_c\rangle _0=(r-1)N$ and a maximal value of $\langle N_c\rangle_{\max}=N-1$ and it provides information about the evolution of our system's synchronizing mechanism. Since the bounding of this value depends on $r$, we introduce a normalized magnitude $\chi$,
\begin{equation}
\chi(T)=\frac{\langle N_c \rangle (T) - \langle N_c\rangle_0}{\langle N_c\rangle_{max}-\langle N_c \rangle_{0}} \quad \quad \langle N_c \rangle = \frac{1}{N} \sum \limits_{i=1}^{N} N_c^i,
\end{equation}
that is a quantification of the \emph{mixing} of our system. When the mixing is minimal ($\langle N_c \rangle = \langle N_c\rangle _0$) $\chi$ is 0. As the system mixes, i.e. the oscillators increase its average number of contacted neighbors, it can grow up to its maximum value $\chi=1$, i.e. $\langle N_c \rangle = \langle N_c\rangle_{\max}$.

%
\begin{figure}[htbp]
\begin{center}
\includegraphics[width=.95\columnwidth]{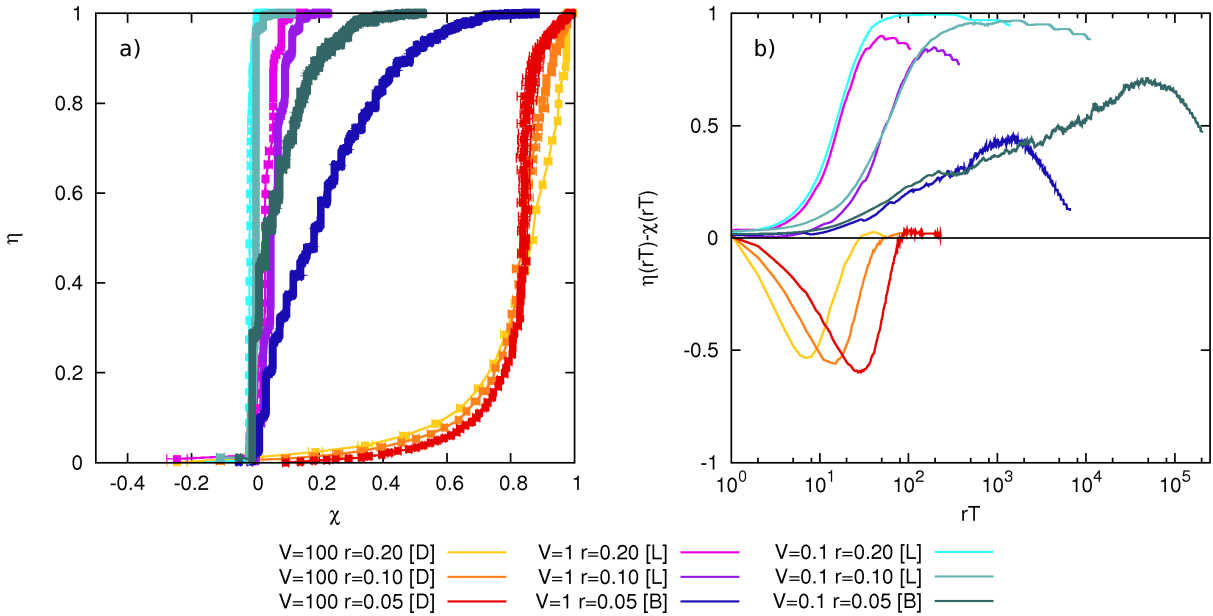}
\caption{(Colors online) Panel $a$): $\eta$ against $\chi$ for several values of $r$ and $V$. Panel $b$): the difference between the two control parameters $(\eta -\chi)$ as a function of $rT$. Letters [D], [L] and [B] stand respectively for "diffusive", "local" and "bounded" regimes. The values of $\eta$ and $\chi$ at each time instant have been calculated averaging over 1000 realizations.}
\label{fig3}
\end{center}
\end{figure}

We have calculated the time evolution of both $\eta$ and $\chi$ for different values of $V\in\{0.1,1,100\}$ and $r\in\{0.05,0.1,0.2\}$. In figure \ref{fig3} one can see the evolution of both parameters measured at the same time instants (and sufficiently averaged over enough realizations) together with the difference between them over time
$\eta(T)-\chi(T)$
that give insight about the topological evolution of our system as it synchronizes. The figures show three clearly distinct patterns.

For high velocities we observe in fig. \ref{fig3}a) a gradual increase of the order parameter and a minor influence on $r$ at fixed $V$, indicating that the synchronization emerges evenly on the system in a global fashion, due to the quickly changing topology of the network (neighbors of a given oscillator change rapidly). This regime (which we call \emph{diffusive}) requires the inter-contact of the majority of the system, but this circumstance is rapidly achieved due to the strong mobility of the agents. In fact, this regime is optimal as far as the synchronization time is concerned, since the interactions are more effective. These conclusions were obtained for populations of Kuramoto oscillators \cite{fkd11}, for which this regime corresponds to the region of validity of the FSA.

In the opposing case, when velocities are small enough, this behavior is completely lost and a step function appears, indicating that the slow mobility of the agents allows them to synchronize locally with their neighbors creating \emph{islands} of synchrony. The sudden increase of the order parameter occurs at a regular pace, fact that points out that whenever the islands are disbanded (change of neighbors), they still transmit the local synchrony to the neighboring groups, mechanism that allows for system synchrony while keeping $\chi$ in small values. The initial height of the steps is dependent on $r$ and decreasing as $\chi(T)$ grows due to the limited range of $\eta(T)$ available states.

Finally, as we decrease $r$ approaching the critical value $r_c$, we observe a transition from a \emph{local} to a \emph{bounded} regime, where the synchronizing time is so long that again allows for the interaction of the majority of the agents among themselves upon synchronization time (due to the bounded nature of the system). In this regime, the range of interaction is very reduced, and so is the size of the clusters, so an agreement between the multiple clusters created (if any) comes after almost all the system has interacted. Consequently, the increasing of $\eta$ with $\chi$ is slower (many small steps, see fig. \ref{fig3}a) ) while the final value $\chi$ becomes larger (fig. \ref{fig3}).

In fig. \ref{fig3}b)
we provide an explicit time evolution of the difference $\eta(T)-\chi(T)$ in order to make the three regimes and the influence of $r$ better identified.

\begin{figure}[ht
Further evidences of this behavior are shown in figure \ref{fig4}. We observe bp]
\begin{center}
\includegraphics[width=.95\columnwidth]{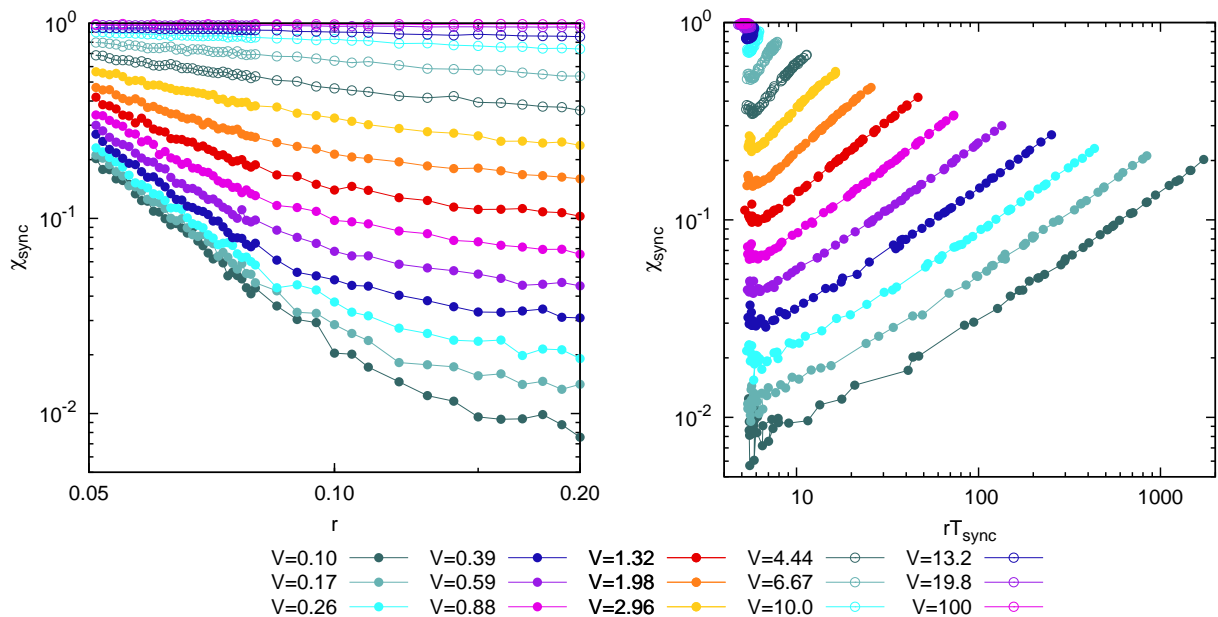}
\caption{(Colors online) On the left: the average value of $\chi$ at the synchronization time ($\chi_{sync}$) as a function of $r$ for several values of $V$. On the right: $\chi_{sync}$ as a function of $rT_{sync}$ for several values of $V$. All the means have been performed over $200$ realizations.}
\label{fig4}
\end{center}
\end{figure}

It is interesting to study the final mixing of our system upon synchronization as shown in figure \ref{fig4}. 
This value $\chi(T=T_{sync})\equiv \chi_{sync}$ together with $T_{sync}$ characterize the evolution of the system towards the synchronized final state. These features depend
on both $r$ and $V$. At fixed velocity the final mixing of the system decreases as the interaction range grows. This is caused by the fact that although an increased $r$ induces more mixing (as oscillators find new neighbors more easily) it also drastically reduces (below the critical values $r_c$) the synchronizing time $T_{sync}$ thus reducing the chances of encounters between different oscillators. Above $r_c$ we observe a saturation of the values as the dependence of $T_{sync}$ in $r$ and $V$ is practically lost. 

Figure \ref{fig4} b) shows a change between the relation pattern of $\chi_{sync}$ and efficiency $rT_{sync}$ ranging from the \emph{diffusive} regime (mixing independent of $rT_{sync}$) to the curves where both the \emph{local} and \emph{bounded} regimes are shown. It is important to note the similarity of the observed shapes (for a wide range of $V$ values) where we find the local phase concentrated around the minimum value of $\chi_{sync}$ that gradually grows in a power law fashion as the performance of the system decreases (it consumes more energy to synchronize).

\begin{figure}
\begin{center}
\includegraphics[width=.95\columnwidth]{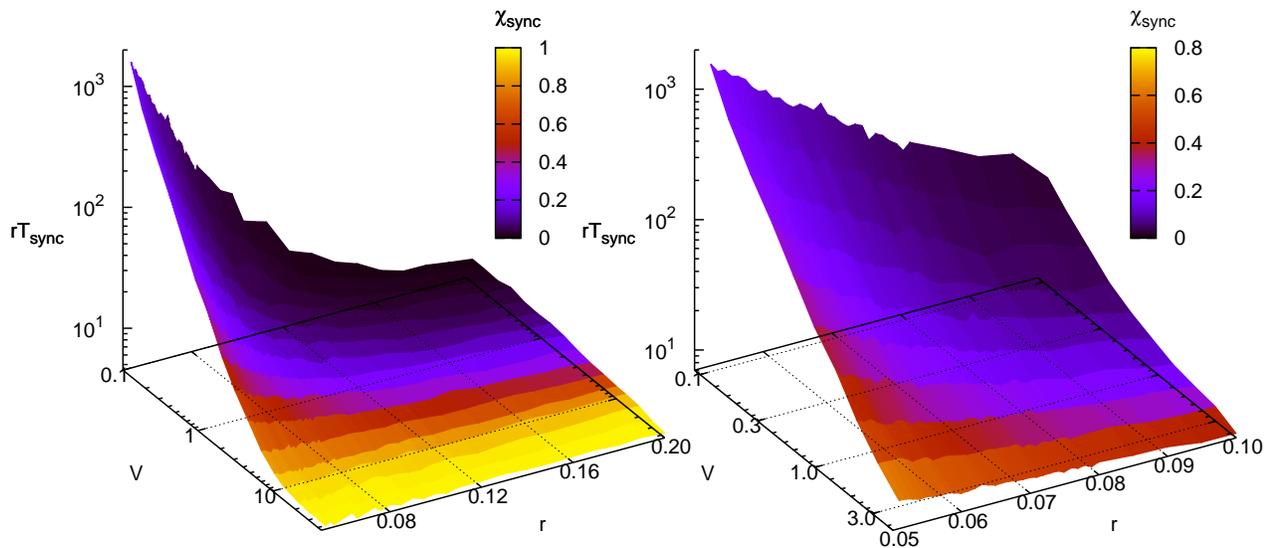}
\caption{(Color online) On the left: The efficiency $rT_{sync}$ as a function of $r$ and $V$. The heat map of $\chi_{sync}$ has been superimposed to the surface $rT_{sync}(r,V)$. On the right: the same quantities, zoomed in the most sensitive region of the parameter space.}
\label{fig5}
\end{center}
\end{figure}

The introduction of the new parameter $\chi_{sync}$ allows us to present a phase diagram of our system relating the overall performance (in terms of efficiency)
with the synchronizing mechanism used (figure \ref{fig5}). 
We identify the \emph{diffusive} regime in the zone of high velocities $V\sim \mathcal{O}(10)$ where both values of $\chi$ and $rT_{sync}$ are almost independent of $r$. This zone falters into the \emph{bounded} one as $V$ and specially $r$ decrease, where both the efficiency and the mixing of the system is reduced. Finally for small enough velocities the local zone is clearly visible with low values of system mixing. From the map one clearly observes that the most beneficial synchronizing mechanisms (in terms of energy consumption) are the diffusive and local ones.

\section{Conclusions}

Following the recent literature on complex systems, one of the hottest topics is the relation between dynamics and topology of interactions. In particular, there many evidences that the patterns of interaction change rapidly with time, thereby completely altering and conditioning the dynamical properties of the system. 

Here we have proposed a framework in which agents move on a plane and are allowed to interact in a pulsing way. Each agent, representing a phase oscillator, moves at a common velocity and changes its internal phase with a common period. When this internal phase reaches a threshold value, the oscillator "fires", thus resetting its own phase as well as  its orientation. At the same time, this oscillator interacts with the neighbors within a certain distance by changing their phases. Notably, this interaction setting makes the system reach a final synchronized state in which all oscillators fire at unison within the same fast scale. 

Keeping all geometrical parameters constant, we analyze the system behavior by changing solely  the velocity of the agents and the interaction range. Notice that the motion of the agents is what ensures that the system will be able to synchronize. If the agents are static, a minimal fixed topology will be required to connect them (what is called a giant component in network terminology).
In contrast, a population of moving agents 
(even when the interaction range is small) will eventually synchronize.

We measure the time needed for the system to synchronize as a function of the two relevant agent parameters, the velocity and the fraction of population they interact with. 
Our model can be applied, for instance, to the field of wire-less communications by introducing the performance, which stands for the total number of signals emitted by the population to reach the synchronized state. Our simulations show that the time required for synchronization ranges widely, depending on the speed of the agents. As expected, the optimal performance is achieved when agents move very fast, irrespective of the interaction range; however, synchronization in systems with low mobilities dramatically depend on the interaction range. 

Finally, we have focused on the mechanisms that the system follows along its path to synchronization. We have introduced a new order parameter, that stands for the fraction of different units each oscillator has interacted with. 
This order parameter complements performance, and a novel phase diagram enables us to identify three different characterizable mechanisms leading to synchronization: i) the diffusive mechanism, where the system very quickly reaches synchronization by means of extremely effective interactions of each oscillator with a large fraction of the population;
ii) the local mechanism, where agents only interact with a small population fraction, but local synchronization is sufficient to lead the system to the globally synchronized state; and finally iii) the bounded mechanism, characterized by very slow motion and short range of interaction, which allow a high degree of  accumulated mixing during a very long synchronizing time. 

The identification of these mechanisms that relate mobility and interaction to synchronization will undoubtedly be crucial in similar models of populations of moving agents. Indeed, understanding these phenomena will help to design optimal protocols to dissect more realistic settings, as well as to predict their behavior, by defining their interaction rules.

\section*{Acknowledgements}
L.P. is supported by the Generalitat de Catalunya through the FI Program.
This work has been supported by the Spanish DGICyT Grant FIS2009-13730, and by the Generalitat de Catalunya 2009SGR00838.

\bibliographystyle{ws-ijbc}
\bibliography{sync_updated}
\end{document}